\documentclass[aps,twocolumn,preprintnumbers,superscriptaddress,amsmath,amssymb,amsfonts]{revtex4-2}
\usepackage{graphicx}  % needed for figures
\usepackage{color}
\usepackage[colorlinks,bookmarks=true,citecolor=blue,linkcolor=blue,urlcolor=blue, breaklinks=true]{hyperref}
\usepackage{bm}
\usepackage{amsmath,amssymb}
\begin{document}

%\preprint{}
\title{Reply to Comment on: Microscopic signatures of an imaginary charge density wave in a kagome metal}

\author{S. Suetsugu}
\affiliation{Department of Applied Physics, The University of Tokyo, Tokyo 113-8656, Japan}
\affiliation{Department of Physics, Kyoto University, Kyoto 606-8502, Japan}
\author{S. Kitagawa}
\affiliation{Department of Physics, Kyoto University, Kyoto 606-8502, Japan}
\author{K. Ishida}
\affiliation{Department of Physics, Kyoto University, Kyoto 606-8502, Japan}
\author{T. Shibauchi}
\affiliation{Department of Advanced Materials Science, University of Tokyo, Chiba, Japan}
\author{Y. Matsuda}
\affiliation{Department of Physics, Kyoto University, Kyoto 606-8502, Japan}
\affiliation{Los Alamos National Laboratory, Los Alamos, New Mexico 87545, USA}

\date{\today}

%\keywords{}


\begin{abstract}
We address a recent Comment [I. Nikolov {\it et al.}, arXiv:2608.13579 (2026)] proposing crystalline mosaicity as an alternative explanation for the asymmetric nuclear magnetic resonance (NMR) spectra reported in our study [S. Suetsugu {\it et al.}, Nat. Phys. {\bf 22}, 1251--1256 (2026)]. We show that this scenario requires substantial temperature- and site-dependent distributions of crystallographic orientations and additional site-dependent distributions of the electric field gradient (EFG) asymmetry parameter, none of which follow from ordinary crystalline mosaicity. We therefore conclude that crystalline mosaicity cannot account for the observed site-selective spectral asymmetry and does not provide an alternative explanation for the central spectroscopic observation underlying our interpretation.
\end{abstract}

\maketitle

A recent Comment \cite{nikolov2026} on our Article "Microscopic signatures of an imaginary charge density wave in a kagome metal" \cite{suetsugu2026microscopic} argues that the asymmetric NMR spectra can be explained by crystalline mosaicity rather than by emergent internal magnetic fields. Here we point out a fundamental inconsistency in the proposed explanation in Ref.\,\citenum{nikolov2026}.

Crystalline mosaicity represents a distribution of crystallographic orientations within a sample. Such a crystallographic distribution should be independent of temperature and common to different nuclear sites in the same crystal. In contrast, the analysis in the Comment \cite{nikolov2026} requires the angular distribution attributed to “mosaicity” to increase from approximately 1.1$^\circ$ at 160\,K to 1.6$^\circ$ at 100\,K for the Sb2 site, and further assumes a substantially larger angular distribution of approximately 2.8$^\circ$ for the Sb1 site. Moreover, the site dependence directly exposes the limitation of the proposed scenario. For the $-1/2\leftrightarrow-3/2$ transition, the observed spectral asymmetry has opposite signs at the Sb1 and Sb2 sites (see Figs.\,3e and S7b in Ref.\,\citenum{suetsugu2026microscopic}). In the calculations of the Comment \cite{nikolov2026}, the angular distribution associated with mosaicity produces the same direction of skewness for this transition at both sites. The opposite asymmetry at Sb1 is reproduced only after introducing additional, site-dependent distributions of the EFG parameters, in particular a substantially larger distribution of the asymmetry parameter $\eta$. Since $\eta$ characterizes the local electronic charge environment of each site and is invariant under any rotation of the crystal, a distribution of $\eta$ cannot originate from orientational disorder. Thus, the Sb1 and Sb2 spectra are not explained by a common mosaicity mechanism. The necessity of introducing substantial temperature-dependent and site-dependent angular distributions, together with an additional site-dependent distribution of $\eta$, demonstrates that crystalline mosaicity cannot account for the observed asymmetric spectra.

While the other points raised in the Comment can also be readily addressed, they do not alter this inconsistency of the proposed mosaicity scenario. We therefore conclude that crystalline mosaicity does not provide an alternative explanation for the central spectroscopic observation underlying our interpretation.

\bibliography{ref.bib}

\end{document}